\documentclass[aps,prb,reprint,onecolumn,notitlepage,superscriptaddress]{revtex4-2}

\usepackage{color}
\usepackage{dcolumn}
\usepackage{amsmath}
\usepackage{amssymb}
\usepackage{mathtools}
\usepackage{graphicx}
\usepackage{latexsym}
\usepackage{amsfonts}
\usepackage{amssymb}
\usepackage{comment}
\usepackage[colorlinks=true,pdfstartview=FitV,linkcolor=black,citecolor=black,urlcolor=blue]{hyperref}

\usepackage{upgreek}
\usepackage{mleftright}

\begin{document}

\title{Stability bounds for the generalized Kadanoff-Baym ansatz in the Holstein dimer}

\author{O. Moreno Segura}
\affiliation{Department of Physics, Nanoscience Center, P.O. Box 35, 40014 University of Jyv{\"a}skyl{\"a}, Finland \looseness=-1}

\author{Y. Pavlyukh}
\affiliation{Institute of Theoretical Physics, Faculty of Fundamental Problems of Technology,
Wroclaw University of Science and Technology, 50-370 Wroclaw, Poland}
\affiliation{Independent Researcher, Karlsruhe, Germany}

\author{R. Tuovinen}
\affiliation{Department of Physics, Nanoscience Center, P.O. Box 35, 40014 University of Jyv{\"a}skyl{\"a}, Finland \looseness=-1}
\email{riku.m.s.tuovinen@jyu.fi}

\begin{abstract}
Predicting real‐time dynamics in correlated systems is demanding: exact two‐time Green's function methods are accurate but often too costly, while the Generalized Kadanoff-Baym Ansatz (GKBA) offers time-linear propagation at the risk of uncontrolled behavior. We examine when and why GKBA fails in a minimal yet informative setting, the Holstein dimer that describes electron-phonon coupling. Using a conserving, fully self‐consistent electron-phonon self‐energy, we map out parameter regions where GKBA dynamics is stable and where it becomes unstable. We trace the onset of these failures to qualitative changes in the model's ground‐state solutions obtained from the full nonequilibrium Green's function theory, thereby providing practical stability bounds for GKBA time evolution. We further show that coupling the dimer to electronic leads can damp and, in part, cure these instabilities. The results supply simple diagnostics and guidelines for reliable GKBA simulations of electron-phonon dynamics.
\end{abstract}

\maketitle

\section{Introduction}\label{sec:intro}

Nonequilibrium dynamics of quantum‐correlated many-body systems presents formidable conceptual and computational challenges. A systematic and conserving description is provided by the nonequilibrium Green's function (NEGF) framework based on the Keldysh-Kadanoff-Baym equations (KBE)~\cite{keldysh1964diagram, kadanoff1962quantum, danielewicz1984quantum, stefanucci2013nonequilibrium, balzer2013nonequilibrium}. However, direct two-time solutions of the KBE entail a cubic scaling with the number of time steps, which renders realistic simulations prohibitively expensive for long propagation times or extended parameter scans.

To reduce the computational complexity, the Generalized Kadanoff-Baym Ansatz (GKBA)~\cite{lipavsky1986generalized} reconstructs two-time Green's functions from single-time quantities, enabling algorithms with linear scaling in the number of time steps. This favorable scaling has spurred a broad range of developments and applications across closed and open systems~\cite{schluenzen2020achieving, karlsson2021fast, pavlyukh2022time-i, pavlyukh2022time-ii, balzer2023accelerating, tuovinen2023time}. At the same time, it has become clear that the GKBA can exhibit uncontrolled behavior in certain regimes: reports include instabilities or unphysical transients in electronic occupations~\cite{hermanns2014hubbard, joost2022dynamically}, pathologies in electron-boson dynamics~\cite{pavlyukh2022time-ii, stefanucci2024semiconductor}, and problematic current evolution in open quantum systems~\cite{kalvova2019beyond, kalvova2023dynamical, pavlyukh2025open, tuovinen2025thermoelectric}. These observations motivate a careful assessment of the dynamical stability of GKBA evolutions.

\begin{figure*}
\centerline{\includegraphics[width=0.5\columnwidth]{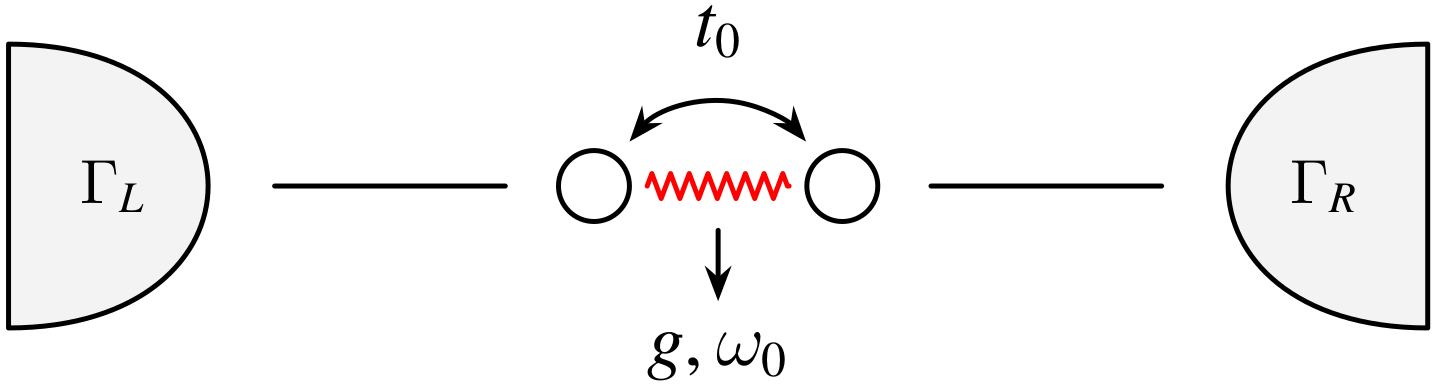}}
\caption{Schematic of the Holstein dimer coupled to leads. The relative phonon mode—characterized by coupling $g$ and frequency $\omega_0$—is shown. $\Gamma_{\alpha}$ ($\alpha=L,R$) denotes the tunneling rate to the leads, and $t_0$ is the hopping term between sites.\label{fig:dimer}}
\end{figure*}

In this work, GKBA stability is analyzed for a paradigmatic electron-phonon setting: the Holstein dimer (see Fig.~\ref{fig:dimer}). Focusing on the fully self-consistent and conserving $GD$ approximation for the electron-phonon self-energy~\cite{sakkinen2015many, sakkinen2015many-ii, karlsson2021fast}, the study maps out parameter regions of stable and unstable GKBA dynamics and relates them to known ``bifurcations'' of the Holstein model's ground-state properties obtained from the full NEGF theory~\cite{sakkinen2015many}. This correspondence provides practical stability bounds for GKBA time evolution in electron-phonon problems. The analysis is then extended to an open-system setting by coupling the dimer to electronic leads within the wide-band limit approximation (WBLA)~\cite{tuovinen2023time}. It is found that the tunneling-induced damping can regularize parts of the unstable GKBA regime in the isolated system, in line with the inclusion of quasi-particle effects in related GKBA applications~\cite{bonitz1996numerical, jahnke1997linear, pal2009conserving, makait2023time}. At the same time, caveats associated with energy conservation in open systems apply~\cite{latini2014charge, bonitz2019ion}, which provide additional context for the present electron-phonon analysis.

The paper is organized as follows. Section~\ref{sec:theory} introduces the Holstein (dimer) model and the governing NEGF+GKBA equations. Section~\ref{sec:results} presents simulations that map stable and unstable regimes across the model's parameter space. Section~\ref{sec:conclusion} summarizes the main findings and outlines avenues for future work.

\section{Theory}\label{sec:theory}
The Hamiltonian for the model of interacting electrons and phonons (see Fig.~\ref{fig:dimer}) studied in this work is the following
\begin{equation}\label{eq:Hd}
    \hat{H}=-t_0(\hat{d}^\dagger_{1} \hat{d}_{2} + \hat{d}^\dagger_{2} \hat{d}_{1}) +\sum_i \omega_i \hat{a}_i^\dagger \hat{a}_i - g \sum_i(\hat{a}_i^\dagger+\hat{a}_i)\hat{n}_{i},
\end{equation}
where $\hat{d}_i$ ($\hat{a}_i$) and $\hat{d}^{\dagger}_i$ ($\hat{a}^{\dagger}_i$) are the fermionic (bosonic) annihilation and creation operators, respectively. $\hat{n}_i=\hat{d}^{\dagger}_i \hat{d}_i$ is the fermionic number operator. The parameter $g$ denotes the interaction strength between electrons and phonons, $t_0$ is the characteristic hopping term, and $\omega_i$ the phonon frequency. To be consistent with the typical formulation in NEGF~\cite{sakkinen2015many, sakkinen2015many-ii}, we rewrite the phononic and interacting Hamiltonian by introducing the displacement operator $\hat{u}_i \equiv \hat{\phi}_{i,1} = (\hat{a}^{\dagger}_{i} + \hat{a}_{i})/\sqrt{2}$, and momentum operator $\hat{p}_i \equiv \hat{\phi}_{i,2} = i(\hat{a}^{\dagger}_{i} - \hat{a}_{i})/\sqrt{2}$,
\begin{align}
    \hat{H}_{\mathrm{p}} (t)= \sum_{\mu \nu} \Omega_{\mu \nu}(t)\hat{\phi}_{\mu} \hat{\phi}_{\nu}, \quad \hat{H}_{\mathrm{e}}(t)=\sum_{ij}h_{ij}(t)\hat{d}^\dagger_{i}\hat{d}_{j}, \quad \hat{H}_{\mathrm{e-p}} (t)= \sum_{\mu,ij} g_{\mu,ij}(t)\hat{d}^\dagger_{i}\hat{d}_{j}\hat{\phi}_{\mu},
\end{align}
where we also introduced the one-electron matrix elements $h_{ij}$. The time dependence of the electron-phonon coupling is due to a switch-on function, which is specified in the next section. The composite index $\mu$ combines the phononic mode label $i$ with $\xi=1,2$ for displacement and momentum, so that $\mu=(i,\xi)$. The Hamiltonian~\eqref{eq:Hd} is recovered by substituting the following relations,
\begin{align}
h =  -t_0
\begin{pmatrix}
0 &1 \\
1 &0
\end{pmatrix}, \quad \Omega_{\mu\nu}\equiv \frac{\omega_i}{2}(\delta_{ij}\delta_{\xi\xi'}+\upalpha_{i\xi,j\xi'}), \quad g_{\mu,jk}\equiv -\sqrt{2}g\delta_{\xi,1}\delta_{i j}\delta_{jk},
\end{align}
where the matrix $\upalpha$ comes from the commutation relations for the $\hat{\phi}$ fields,
\begin{equation}
[\hat{\phi}_{\mu},\hat{\phi}_{\nu}] = \upalpha_{\mu \nu},~~~~ \upalpha_{\mu \nu} =  \delta_{ij}
\begin{pmatrix}
0 &i \\
-i &0
\end{pmatrix}_{\xi \xi'}.
\end{equation}

The effect of electronic leads is modeled by
\begin{align}
\hat{H}_{\mathrm{lead}}&=\sum_{k\alpha} \epsilon_{k\alpha}\hat{d}_{k\alpha}^\dagger\hat{d}_{k\alpha}, \\
\hat{H}_{\mathrm{tunneling}}(t)&=\sum_{ik\alpha}\left[T_{k\alpha,i}(t)\hat{d}_{k\alpha}^\dagger\hat{d}_i + \mathrm{H.c.}\right],
\end{align}
where $\epsilon_{k\alpha}$ is the energy dispersion relation for the $k$-th state in the $\alpha$-th lead, and $T_{k\alpha,i}$ is the tunneling matrix element between the lead and dimer states. The latter object is time-dependent, similar to $g$, with a switch-on function specified below.

In the context of NEGF, we are interested in the one-particle Green's functions (GFs), and as we work here under the fully self-consistent Born approximation, i.e., the so-called $GD$ approximation~\cite{karlsson2021fast,pavlyukh2022time-i}, we take into account the electronic and phononic GFs, with lesser and greater components
\begin{align}
G^{<}_{ij}(t,t')=i\langle \hat{d}^\dagger_{j}(t') \hat{d}_{i}(t) \rangle, \quad G^{>}_{ij}(t,t')=-i\langle \hat{d}_{i}(t) \hat{d}^\dagger_{j}(t') \rangle, \quad D^{<}_{\mu  \nu}(t,t')=D^{>}_{  \nu\mu}(t',t)=-i\langle \Delta \hat{\phi}_{\nu}(t')  \Delta\hat{\phi}_{\mu}(t) \rangle,
\end{align} 
where $\Delta\hat{\phi}_{\nu}(t)\equiv \hat{\phi}_{\nu}(t) - \langle \hat{\phi}_{\nu}(t) \rangle$ is the fluctuation operator and $\langle \hat{\phi}_{\nu}(t) \rangle = \phi_{\nu}(t)$ is the expectation value of the phononic field. These two correlators satisfy their corresponding Kadanoff-Baym equations in matrix representation~\cite{karlsson2021fast}:
\begin{align}
[i \partial_t- h^{\mathrm{e}}(t)]G^{\lessgtr}(t,t') & =[\Sigma^{\lessgtr}\cdot G^A + \Sigma^{R}\cdot G^{\lessgtr}](t,t'), \label{eq:KBE-g} \\
[i\partial_{t} - h^{\mathrm{p}}(t)]D^{\lessgtr}(t,t') & =\upalpha \left[\Pi^{\lessgtr}\cdot D^A + \Pi^{R}\cdot D^{\lessgtr}\right] (t,t'),\label{eq:KBE-d}
\end{align}
where superscripts $A$ and $R$ stand for the advanced and retarded components, respectively, and $[a \cdot b](t,t')\equiv\int d\bar{t}~a(t,\bar{t})b(\bar{t},t')$ denotes a real-time convolution. The electronic self-energy $\Sigma=\Sigma_{\mathrm{c}}+\Sigma_{\mathrm{em}}$ consists of the correlation $\Sigma_{\mathrm{c}}$ and embedding $\Sigma_{\mathrm{em}}$ parts. The former accounts for the electron-phonon interaction, while the latter for the system-leads coupling. The electronic Hamiltonian $h^{\mathrm{e}}(t)$ includes the time-local mean-field contribution, so it is the one-body Hamiltonian and the phononic potential, $h^{\mathrm{e}}_{ij}(t)= h_{ij}(t)+ \sum_{\mu}g_{\mu,ij}(t)\phi_{\mu}(t)$. The term $\Pi$ is the phononic self-energy, and $h^{\mathrm{p}} \equiv \upalpha(\Omega +\Omega^T)$ the effective phononic Hamiltonian. Within the $GD$ scheme, the phonon self-energy corresponds to the electron-hole bubble (random-phase approximation) without vertex corrections~\cite{keating1968dielectric, vogl1976microscopic, giustino2017electron}, which we adopt throughout this work.

The equation of motion (EOM) for the electronic density matrix $\rho_{\mathrm{e}}(t)=-iG^{<}(t,t)$ and the phononic density matrix $\rho_{\mathrm{p}}(t) = iD^{<}(t,t)$ can be obtained from Eqs.~\eqref{eq:KBE-g}, \eqref{eq:KBE-d} and their adjoints, and taking the limit $t=t'$,
\begin{subequations}
\begin{align}
    i\frac{d}{dt}\rho_{\mathrm{e}}(t) & = \left[h^{\mathrm{e}}(t)\rho_{\mathrm{e}}(t) -iI_{\mathrm{c}}(t) -iI_{\mathrm{em}}(t)\right]- \mathrm{H.c.}, \label{eq:EOM_rhos1}\\
    i\frac{d}{dt} \rho_{\mathrm{p}}(t) & = \left[h^{\mathrm{p}}(t)\rho_{\mathrm{p}}(t)  +iI_{\mathrm{p}}(t)\right] -\mathrm{H.c.}, \label{eq:EOM_rhos2}
\end{align}
\end{subequations}
where $I_\mathrm{c}$, $I_{\mathrm{em}}$, and $I_{\mathrm{p}}$ are the collision integrals, which come from the convolution of the corresponding self-energies and the GFs. To make the collision integrals functionals of the density matrices $\rho_{\mathrm{e}}(t)$ and $\rho_{\mathrm{p}}(t)$ we employed the GKBA~\cite{karlsson2021fast}
\begin{align}
    G^{\lessgtr}(t,t') & = \mp\left[G^R(t,t')\rho_{\mathrm{e}}^{\lessgtr}(t') -\rho_{\mathrm{e}}^{\lessgtr}(t)G^A(t,t') \right], \\
    D^{\lessgtr}(t,t') & = D^R(t,t')\upalpha\rho_{\mathrm{p}}^{\lessgtr}(t') -\rho_{\mathrm{p}}^{\lessgtr}(t)\upalpha D^A(t,t'),
\end{align}
and we approximate the retarded and advanced GFs at the mean-field level
\begin{align}
    G^R(t,t') & = -i\theta(t-t')\mathcal{T}\left[ \exp\left( -i\int^{t}_{t'}d\bar{t}[h^{\mathrm{e}}(\bar{t})-\frac{i}{2}\Gamma(\bar{t})]\right)\right],\label{eq:propag-g}\\
    D^{R}(t,t') & = -i\upalpha\theta(t-t')\mathcal{T}\left[ \exp\left( -i\int^{t}_{t'}d\bar{t}~h^{\mathrm{p}}(\bar{t})\right)\right],\label{eq:propag-d}
\end{align}
with $\Gamma(t)\equiv\sum_{\alpha}s_{\alpha}^2(t)\Gamma_{\alpha}$, where $s_{\alpha}(t)$ is the switch-on function for the coupling between the lead $\alpha$ and the system, and $\Gamma_{\alpha}$ is the line-width matrix of lead $\alpha$ at the WBLA level~\cite{tuovinen2023time}:
\begin{equation}
\Gamma_{\alpha,ij}=2\pi\sum_k T_{i,k\alpha}\delta(\mu-\epsilon_{k\alpha})T_{k\alpha,j},
\end{equation}
where $\mu$ is the chemical potential.
For $\Gamma=0$, $G^R$ is reduced to the propagator for closed systems, which is used when the leads are disconnected ($\Sigma_{\mathrm{em}}=0$). 

The time-linear method is constructed by first rewriting the collision integrals $I_c$ and $I_\mathrm{p}$ in terms of a higher-order correlator $\mathcal{G}^{\mathrm{e-p}}$ (approximated here at the $GD$ level), and then deriving its EOM as introduced in Refs.~\citenum{karlsson2021fast, pavlyukh2022time-i}. A similar procedure is applied to $I_\mathrm{em}$ by defining an embedding correlator $\mathcal{G}^{\mathrm{em}}$ under the wide-band limit approximation and finding its EOM, as developed in Ref.~\citenum{tuovinen2023time}. Based on these considerations, we rewrite Eq.~\eqref{eq:EOM_rhos1} for $\rho_{\mathrm{e}}(t)$ as
\begin{align}
    i\frac{d}{dt}\rho_{\mathrm{e},ij}(t) = \left([h^{\mathrm{e}}_{\mathrm{eff}}(t)\rho_{\mathrm{e}}(t)]_{ij} +\frac{i}{4}\Gamma_{ij}(t) + i \sum_{l\alpha}s_{\alpha}(t)\frac{\eta_l}{\beta}\Gamma_{\alpha,ij}\mathcal{G}^{\mathrm{em}}_{l\alpha}(t) 
     + \sum_{\mu,k} g_{\mu,ik}(t)\mathcal{G}^{\mathrm{e-p}}_{\mu,kj}(t) \right)- \mathrm{H.c.},\label{eq:newEOM_rhosel}
\end{align}  
where $h^{\mathrm{e}}_{\mathrm{eff}}(t)\equiv h^{\mathrm{e}}(t)-i\,\Gamma(t)/2$ is the effective Hamiltonian, $\eta_l$ arise from the pole expansion of the Fermi function~\cite{hu2010communication}, and $\beta$ denotes the inverse temperature. Similarly, Eq.~\eqref{eq:EOM_rhos2} for $\rho_{\mathrm{p}}(t)$ can be rewritten as 
\begin{align}    
    i\frac{d}{dt} \rho_{\mathrm{p},\mu\nu}(t) = \left([h^{\mathrm{p}}(t)\rho_{\mathrm{p}}(t)]_{\mu\nu}  + \sum_{ij} \bar{g}_{\mu,ij}(t)\mathcal{G}^{\mathrm{e-p}}_{\nu,ji}(t)\right) -\mathrm{H.c.}\label{eq:newEOM_rhosph},
\end{align}
where $\bar{g}_{\mu,ij}=\sum_{\nu}\upalpha_{\mu\nu}g_{\nu,ij}$. Equations~\eqref{eq:newEOM_rhosel} and~\eqref{eq:newEOM_rhosph} must be coupled to the EOM for the higher-order electron–phonon correlator
\begin{align}
i\frac{d}{dt} \mathcal{G}^{\mathrm{e-p}}_{\mu,ij}(t) = -\Psi_{\mu,ij}(t) + \sum_{\sigma}h_{\mu \sigma}^{\mathrm{p}}(t) \mathcal{G}^{\mathrm{e-p}}_{\sigma,ij}(t) +\sum_k [h^{\mathrm{e}}_{\mathrm{eff}, \, ik}(t)\mathcal{G}^{\mathrm{e-p}}_{\mu,kj}(t)-\mathcal{G}^{\mathrm{e-p}}_{\mu,ik}(t)h^{\mathrm{e}}_{\mathrm{eff},\, kj}(t)],\label{eq:elph}
\end{align}
where $\Psi_{\mu,ij}(t) = \Psi^>_{\mu,ij}(t)-\Psi^<_{\mu,ij}(t)$ with $\Psi^{\lessgtr}_{\mu,ij}(t)=\sum_{\nu,sq}\rho_{\mathrm{p},\mu\nu}^{\lessgtr}(t)\rho_{\mathrm{e},is}^{\lessgtr}(t)g_{\nu,sq}(t)\rho^{\gtrless}_{\mathrm{e},qj}(t)$. Equation~\eqref{eq:newEOM_rhosel} is also coupled to the EOM of the embedding correlator 
\begin{align}\label{eq:emb}
i\frac{d}{dt}\mathcal{G}^{\mathrm{em}}_{l\alpha}(t)=-s_{\alpha}(t)-\mathcal{G}^{\mathrm{em}}_{l\alpha}(t)\left(h^{\mathrm{e}\,\dagger}_{\mathrm{eff}}(t)-V_{\alpha}(t)-\mu +i\frac{\zeta_l}{\beta}\right),
\end{align}
where $V_\alpha(t)$ is the bias-voltage profile and $\zeta_l$ result from the pole expansion of the Fermi function~\cite{hu2010communication}.
Additionally, the evolution of the expectation values $\phi_{\mu}$, which enter $h^{\mathrm{e}}$, must be taken into account
\begin{equation}\label{eq:field}
    i\frac{d}{dt} \phi_{\mu}(t)=\sum_{\nu}h^{\mathrm{p}}_{\mu\nu}(t)\phi_{\nu}(t)+\sum_{ij,\sigma}\upalpha_{\mu\sigma}g_{\sigma,ij}(t)\rho_{\mathrm{e},ji}(t).
\end{equation}
This set of coupled ordinary differential equations, \eqref{eq:newEOM_rhosel}, \eqref{eq:newEOM_rhosph}, \eqref{eq:elph}, \eqref{eq:emb}, and~\eqref{eq:field}, defines the framework used to describe the system. Within this formalism, we evaluate observables such as the electronic natural occupation (i.e., the eigenvalues of $\rho_{\mathrm{e}}$), the phonon number, and the interaction energies. The phonon number is given by 
\begin{equation}
    n_{\mathrm{p,i}}(t)=\frac{1}{2}\left( \sum_{\xi} \left(\rho_{\mathrm{p},i\xi,i\xi}(t) + \left[\phi_{i\xi}(t)\right]^2 \right)-1 \right).
\end{equation}
The total energy is computed from the mean-field contributions of the electronic, phononic, and electron–phonon terms, together with the correlation energies, as follows
\begin{equation}
    E=E_{\mathrm{e}}(t)+E_{\mathrm{p}}(t)+E_{\mathrm{e-p}}(t)+E_{\mathrm{e,c}}(t)+E_{\mathrm{p,c}}(t),
\end{equation}
where each term is calculated according to 
\begin{align}
    E_{\mathrm{e}}(t) & = \mathrm{Tr}\left[ h(t)\rho_{\mathrm{e}}(t)\right]\\
    E_{\mathrm{p}}(t) & = \mathrm{Tr}\left[ \Omega(t)\Lambda(t)\right],\\
    E_{\mathrm{e-p}}(t) & = \mathrm{Tr}\left[ h_{\mathrm{e-p}}(t)\rho_{\mathrm{e}}(t)\right],\\
    E_{\mathrm{e,c}}(t) & = -\frac{i}{4}\mathrm{Tr}\left[ I_\mathrm{c}(t) - I^{\dagger}_\mathrm{c}(t)\right],\\
    E_{\mathrm{p,c}}(t) & = \frac{i}{4}\mathrm{Tr}\left[ \upalpha I_{\mathrm{p}}(t) +(\upalpha I_{\mathrm{p}}(t))^{T}\right],
\end{align}
where $h_{\mathrm{e-p},ij}(t)=\sum_{\mu} g_{\mu,ij}(t)\phi_{\mu}(t)$ and $\Lambda_{\mu\nu}=\rho_{\mathrm{p},\mu\nu}(t)+\phi_{\mu}(t)\phi_{\nu}(t)$. The correlation energies are expressed in terms of the collision integrals for readability; however, they are evaluated using the correlators $\mathcal{G}^{\mathrm{em}}$ and $\mathcal{G}^{\mathrm{e-p}}$, approximated as described above. All our calculations following this scheme are performed with the CHEERS code~\cite{pavlyukh2025cheers}. We use a fixed time step $0.01$ (in units of inverse hopping $t_0^{-1}$) and a fourth-order Runge-Kutta integrator controlling the numerical error within $10^{-9}$.

\section{Results}\label{sec:results}

The system defined in Eq.~\eqref{eq:Hd} is characterized by two dimensionless parameters, the adiabatic ratio $\gamma$ and the effective interaction $\lambda$, given by
\begin{equation}
    \gamma \equiv \frac{\omega_0}{t_0},~~~\lambda \equiv \frac{g^2}{\omega_0 t_0}.
\end{equation}
In the following, the hopping term is set as the energy unit ($t_0=1$); hence $\gamma$ is determined solely by the phonon frequency $\omega_0$. Times are thus expressed in units of inverse hoppings. In addition, the electronic particle number is fixed to one.

\begin{figure*}
\centerline{\includegraphics[width=0.5\columnwidth]{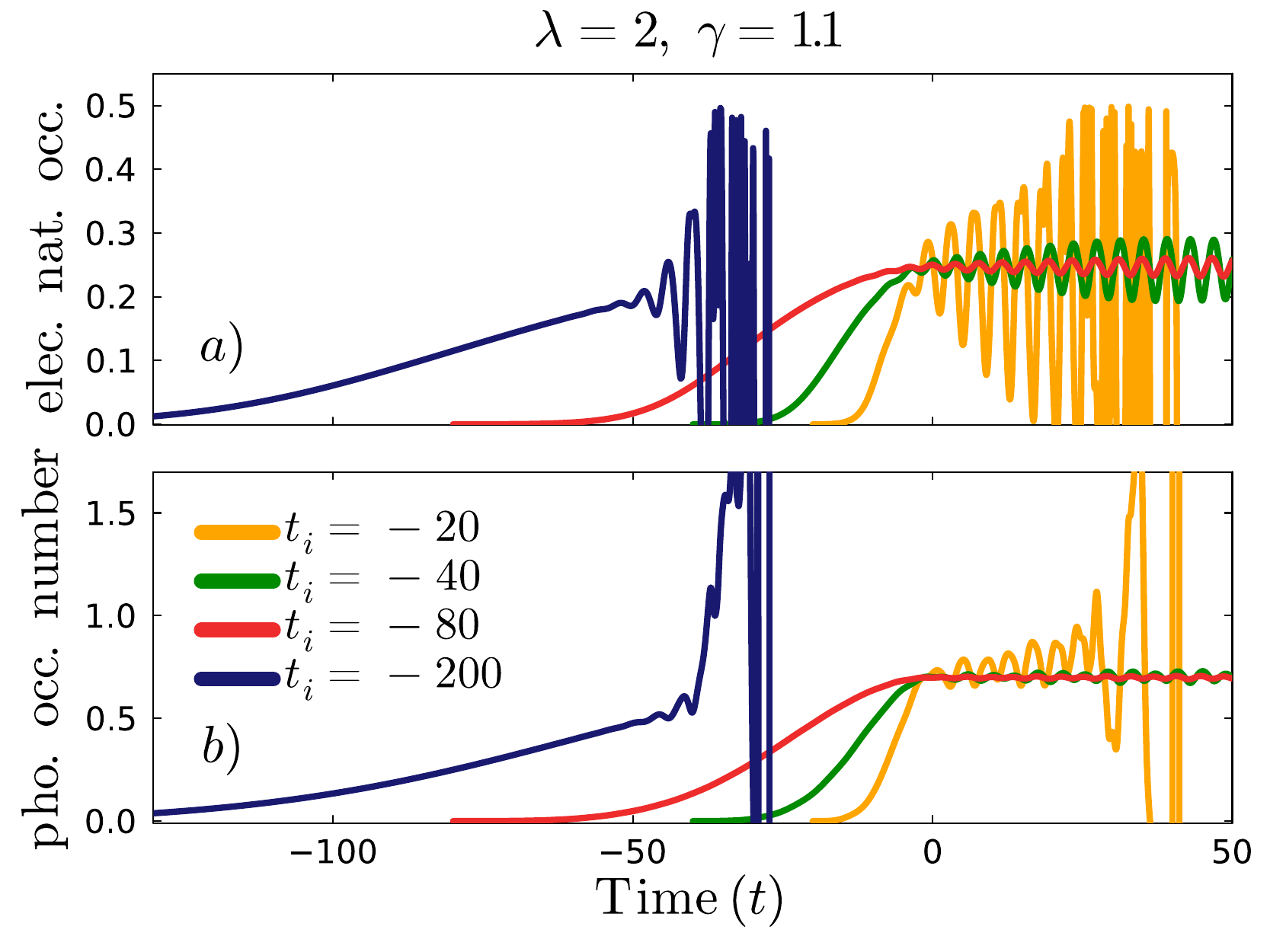}}
\caption{Electronic natural occupation number (a) and phononic occupation number (b) as a function of time for different switching times.\label{fig:populations}}
\end{figure*}

To start our analysis, we study the dynamics in the isolated system ($\Sigma_{\mathrm{em}}=0$) through the electronic natural occupation and the phononic occupation number. An important feature we control in our simulations is the time-dependent coupling, where starting from the uncoupled electron-phonon system, it follows the switching protocol $g(t)=g\cos^{2}\left(\frac{\pi t}{2 t_i}\right)$ for $t<0$. For $t\geq 0$, $g(t)=g$. Therefore, we present different cases by varying the initial time $t_i$ of the simulation, and all cases are propagated until $t_f=50$. An example of a time-dependent calculation is shown in Fig.~\ref{fig:populations}, discussed in detail below.

Figure~\ref{fig:populations} shows a representative case, for $\lambda=2$ and $\gamma=1.1$, that can be either stable or unstable depending on the initial time $t_i$. These cases are informative because they capture the typical behavior of both regimes. Here, stability is defined operationally in an observational sense: we analyze the time evolution within a specified time window and interpret the occurrence of numerical instabilities as evidence of unstable dynamics. For the identified stable cases in our time window ($t_f=50$), $t_i=-40$ and $t_i=-80$, the occupations approach a finite value at $t=0$ and subsequently exhibit small oscillations around it. We have verified (not shown) that these oscillations remain stable even for prolonged time evolutions up to $t_f=150$. For the unstable situations, two different scenarios occur, but both are clearly related to the divergence in the phononic occupation number. For $t_i=-20$, the dynamics break down at positive times after a period of strong oscillations in the electronic natural occupation. In contrast, for $t_i=-200$ the simulation breaks down already at negative times (during the switching stage), where a shorter period of oscillations in the natural occupation is followed by a sudden increase in the phononic occupation that triggers the numerical failure. This behavior is consistent with the system spending an extended time in the critical interaction regime during long switching protocols, allowing unstable modes to develop already during the preparation stage.

\begin{figure*}
\centerline{\includegraphics[width=0.5\columnwidth]{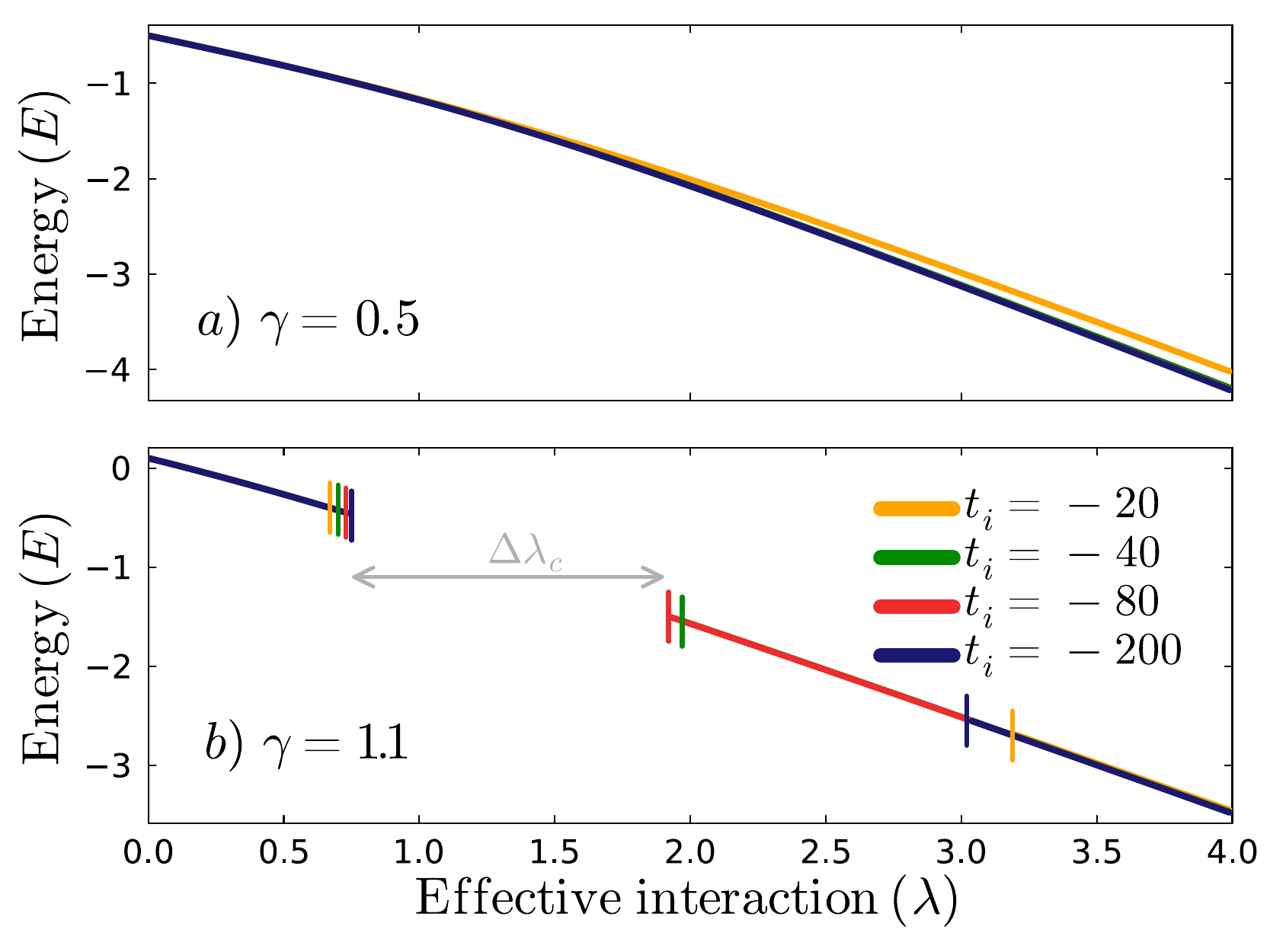}}
\caption{Total energy as a function of the effective interaction for different switching times. The region in panel (b) where the curves show discontinuities—blank gaps—corresponds to instabilities. The difference between the upper and lower critical values is indicated by $\Delta \lambda_c$ for the case $t_i=-80$. The higher adiabatic ratios $\gamma$ in panel (b) correspond to the temporal evolutions of Fig.~\ref{fig:populations}. In panel (a), all curves except for $t_i=-20$ are essentially superimposed. In panel (b), the vertical segments signify the boundaries of the discontinuity regions.}
\label{fig:totalenergy}
\end{figure*}

We then calculate the total energy as a function of the effective interaction for several values of the adiabatic ratio. Figure~\ref{fig:totalenergy} shows two representative cases that illustrate the regimes $\gamma<0.9$ and $\gamma>0.9$, where the total energy is taken to be its value at $t_f$. The case $\gamma=0.5$ (Fig.~\ref{fig:totalenergy}a) is completely stable, and the energy behavior is in agreement with the results presented in Ref.~\citenum{sakkinen2015many}; the two distinct lines arise because $t_i=-20$ is insufficient to reach a steady state at large $\lambda$, the other cases converge to the same value and therefore overlap. By contrast, $\gamma=1.1$ (Fig.~\ref{fig:totalenergy}b) corresponds to a typical unstable case, where the lower critical value of the effective interaction is found around $\lambda \approx 0.75$. The extent of the unstable range depends on the initial time: very fast ($t_i=-20$) and very slow ($t_i=-200$) switching reduce stability (overly abrupt vs. prolonged propagation). The situations for $t_i=-40$ and $t_i=-80$ are approximately similar (the two curves almost overlap throughout), with the latter resulting in the most stable case.

\begin{figure*}
\centerline{\includegraphics[width=0.5\columnwidth]{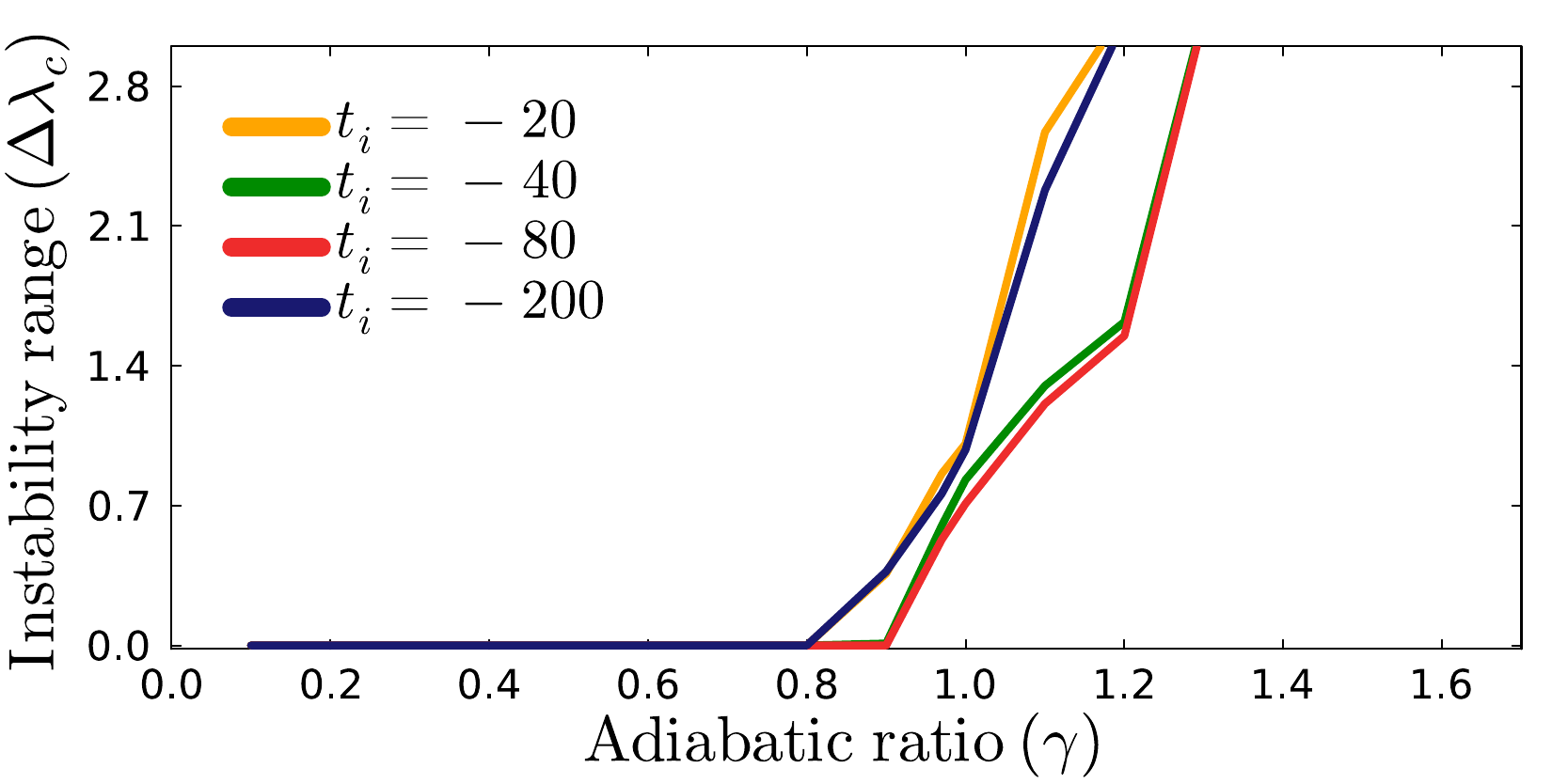}}
\caption{Unstable range of the effective interaction as a function of the adiabatic ratio for several switching times. Instability is signaled by the curves deviating from zero.\label{fig:critical}}
\end{figure*}

In Fig.~\ref{fig:critical}, we show the instability range $\Delta \lambda_c$ of the effective interaction—defined as the difference between the lower and upper critical values of $\lambda$  (cf.~Fig.~\ref{fig:totalenergy}b)—as a function of the adiabatic ratio, thereby characterizing the emergent instabilities in the parameter space $(\lambda,\gamma)$. For all $\gamma\leq 0.8$, the four cases remain stable within the explored range $0\leq\lambda\leq 4$. Conversely, once the adiabatic ratio exceeds approximately $0.9$, instabilities appear in all cases and grow exponentially fast. Importantly, when the effective interaction is smaller than approximately $0.75$, the simulations remain stable for all adiabatic ratios studied here ($\gamma \le 4$). In this weak-coupling regime, correlation effects remain sufficiently small that the GKBA propagation remains consistent with the underlying quasiparticle description, which suppresses the onset of unstable dynamics.

Our results for the unstable range of the effective interaction resemble the behavior of the critical values in $\lambda$ associated with the emergence of an asymmetric solution reported in Ref.~\citenum{sakkinen2015many}, based on the full KBE theory. As shown by Säkkinen \emph{et al.}~\cite{sakkinen2015many}, at the mean-field level, this solution appears as a bifurcation produced by symmetry breaking, independent of the adiabatic ratio, whereas for the fully self-consistent $GD$ approximation, the asymmetric solution emerges discontinuously and has a strong dependence on the adiabatic ratio. In our case, although we employ the $GD$ approach, we close the GKBA with mean-field propagators, Eqs.~\eqref{eq:propag-g} and~\eqref{eq:propag-d}, which leads to an interplay of both scenarios. First, we find that the lower unstable value of $\lambda$ is approximately constant ($\approx 0.75$); and second, the upper unstable value of $\lambda$ is strongly dependent on $\gamma$. Therefore, this is an evident connection between the existence of an asymmetric solution to the instabilities in the GKBA for this model. We emphasize that this interpretation is intended as a physically motivated perspective rather than a formal proof, providing qualitative insight into why the HF-GKBA inherits features of both mean-field and fully self-consistent KBE dynamics.

\begin{figure}[t]
\centering
\includegraphics[width=0.5\textwidth]{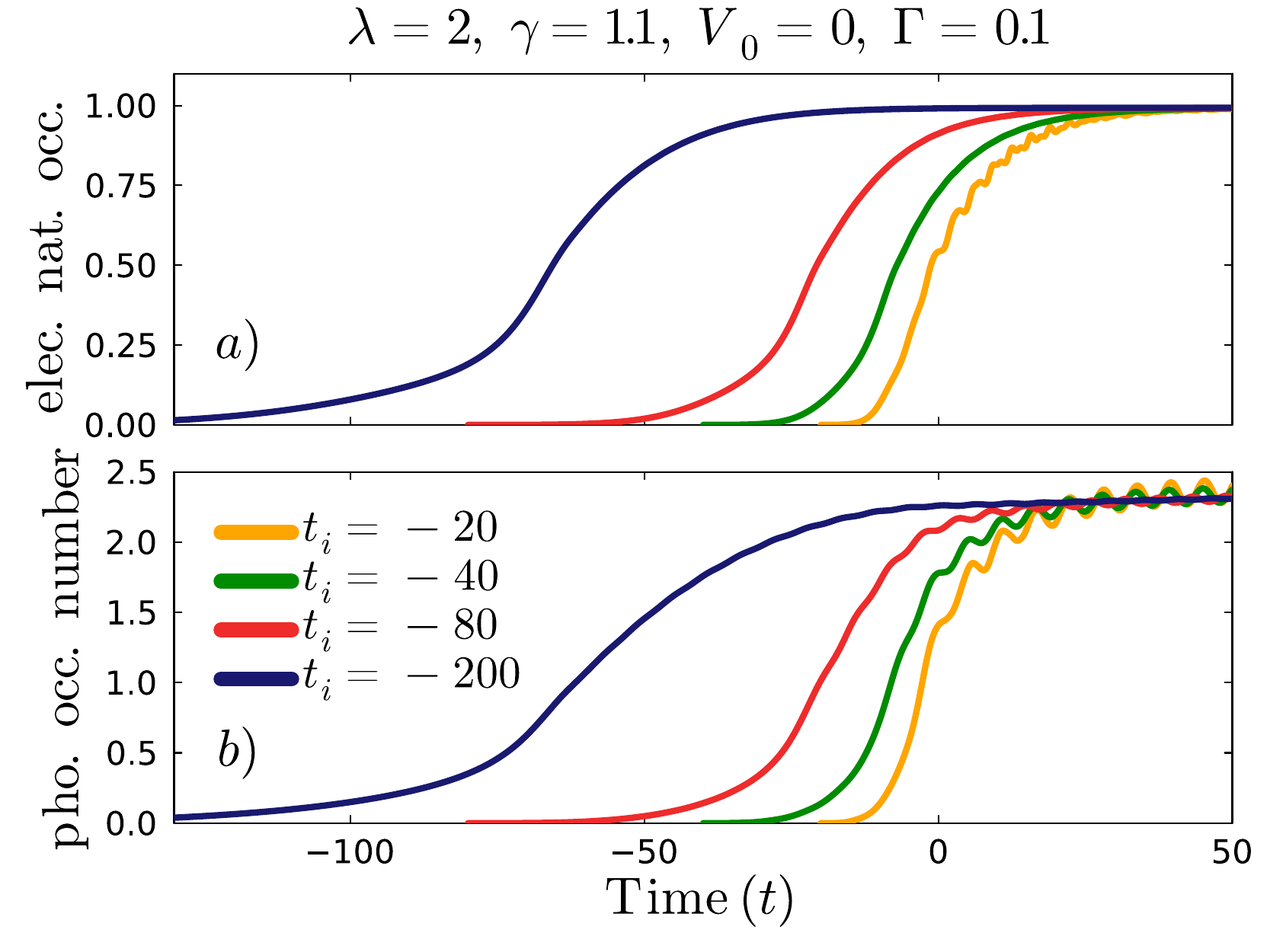}
\caption{Electronic natural occupation number and phononic occupation number as a function of time for the system connected to leads. Results are shown for several switching times.}
\label{fig:leads}
\end{figure}

Previous work~\cite{pavlyukh2024nonequilibrium} has shown that adding a velocity damping term stabilizes correlated electron dynamics in a Hubbard dimer. Here, we investigate whether coupling the Holstein dimer system to electronic leads, described within the WBLA, can achieve a similar stabilization of the electron-phonon dynamics. Therefore, in this part we open the system ($\Sigma_{\mathrm{em}}\neq 0$), and the embedding correlator in Eq.~\eqref{eq:emb} becomes nonzero. This is done by connecting one lead labeled $\alpha=L$ to the site on the left and the other labeled $\alpha=R$ to the site on the right (see Fig.~\ref{fig:dimer}). 
The chemical potential is fixed at the middle of the energy gap of the uncontacted system.
The inverse temperature parameter is fixed to $\beta=10$. As shown in Fig.~\ref{fig:leads}, applying the embedding scheme to the case discussed above effectively removes the instabilities. The case presented here is for zero voltage bias [$V_\alpha=0$ in Eq.~\eqref{eq:emb}] and a line-width matrix with norm $\Gamma=0.1$. 

The effective suppression of spurious oscillations—and the resulting divergences—is evident from the explicit form of the electronic propagator $G^R(t,t')$ in Eq.~\eqref{eq:propag-g}, and the role of $\Gamma$, which produces an exponential decay. In this sense, it is intuitive to follow the system’s dependence on $\Gamma$ as a control parameter. It is found that for more complicated cases (stronger instabilities), such as $\gamma>1.4$, the $\Gamma$ required to stabilize the simulation must be larger. The simulations for these open systems were propagated up to $t_f=8000$ and remained stable. However, although this provides an alternative to control the instabilities, we identify an unphysical increase in the phonon number: unlike the electronic natural occupation, which reaches a steady finite value, the phononic occupation continues to grow slowly. 
We have tested (not shown) that the slope of the phonon occupation number increases with $\Gamma$. This could be related to the larger broadening of the electronic levels (reducing Pauli blocking) increasing the energy accumulation in the phonon modes. In our model, the phonon sector lacks its own dissipation channel, but we will defer more thorough investigations of this effect to future work.
We also caution that once $\Gamma$ becomes comparable to the intrinsic energy scales ($t_0$, $\omega_0$, $g$) of the dimer, the open and isolated problems are no longer directly comparable: increasing $\Gamma$ does not merely regularize the dynamics but opens a lead-mediated scattering channel and enables net energy exchange with the leads, thus modifying the physical evolution.

\section{Conclusion}\label{sec:conclusion}

This work has examined the dynamical stability of the GKBA for a minimal yet physically meaningful electron-phonon problem, the Holstein dimer. Owing to its simplicity and analytical transparency, the dimer enables a controlled study of GKBA evolutions. Within the conserving $GD$ approximation, regions of stable and unstable GKBA dynamics were identified. The onset of instability in time evolution coincides with ground-state ``bifurcations'', which were previously characterized using the full NEGF theory for the same model, as qualitative changes in the stationary solutions~\cite{sakkinen2015many}. This correspondence yields practical bounds for reliable GKBA simulations of electron-phonon dynamics.

Coupling the dimer to WBLA leads was shown to mitigate parts of the unstable regime through dissipative damping, echoing quasi-particle effects noted in GKBA transport applications~\cite{tuovinen2020comparing, tuovinen2021electronic}. While such open-system regularization can be beneficial, it necessarily introduces energy exchange with the environment and must be assessed with the conserving properties of the chosen propagators in mind~\cite{bonitz1996numerical, latini2014charge, tuovinen2023time}. We caution that the observed instabilities are not merely prolonged transients: they arise from a breakdown of the GKBA reconstruction in restricted phonon-energy windows of the isolated Holstein dimer, whereas long-lived transient oscillations can persist as a physical strong-polaronic feature even under numerically exact time evolution~\cite{albrecht2013long, souto2015transient, souto2018transient}.

Altogether, our results map stability domains for GKBA in a canonical electron-phonon setting and suggest concrete diagnostics for anticipating instabilities. Future studies can extend these ideas to larger lattices and nonequilibrium transport setups with structured, non-wide band reservoirs~\cite{pavlyukh2025open, tuovinen2025thermoelectric}, thereby testing the generality of the identified stability bounds and refining GKBA formulations for robust electron-boson dynamics.

\acknowledgments

R.T. and Y.P. wish to acknowledge fruitful discussions over the years on nonequilibrium Green's function theory with Prof. Michael Bonitz, whose 65th birthday is celebrated in this Special Issue. O.M.S. and R.T. acknowledge the financial support of the Jane and Aatos Erkko Foundation (Project EffQSim). R.T. further acknowledges the Research Council of Finland through the Finnish Quantum Flagship (Project No. 359240).


\begin{thebibliography}{38}%
\makeatletter
\providecommand \@ifxundefined [1]{%
 \@ifx{#1\undefined}
}%
\providecommand \@ifnum [1]{%
 \ifnum #1\expandafter \@firstoftwo
 \else \expandafter \@secondoftwo
 \fi
}%
\providecommand \@ifx [1]{%
 \ifx #1\expandafter \@firstoftwo
 \else \expandafter \@secondoftwo
 \fi
}%
\providecommand \natexlab [1]{#1}%
\providecommand \enquote  [1]{``#1''}%
\providecommand \bibnamefont  [1]{#1}%
\providecommand \bibfnamefont [1]{#1}%
\providecommand \citenamefont [1]{#1}%
\providecommand \href@noop [0]{\@secondoftwo}%
\providecommand \href [0]{\begingroup \@sanitize@url \@href}%
\providecommand \@href[1]{\@@startlink{#1}\@@href}%
\providecommand \@@href[1]{\endgroup#1\@@endlink}%
\providecommand \@sanitize@url [0]{\catcode `\\12\catcode `\$12\catcode
  `\&12\catcode `\#12\catcode `\^12\catcode `\_12\catcode `\%12\relax}%
\providecommand \@@startlink[1]{}%
\providecommand \@@endlink[0]{}%
\providecommand \url  [0]{\begingroup\@sanitize@url \@url }%
\providecommand \@url [1]{\endgroup\@href {#1}{\urlprefix }}%
\providecommand \urlprefix  [0]{URL }%
\providecommand \Eprint [0]{\href }%
\providecommand \doibase [0]{https://doi.org/}%
\providecommand \selectlanguage [0]{\@gobble}%
\providecommand \bibinfo  [0]{\@secondoftwo}%
\providecommand \bibfield  [0]{\@secondoftwo}%
\providecommand \translation [1]{[#1]}%
\providecommand \BibitemOpen [0]{}%
\providecommand \bibitemStop [0]{}%
\providecommand \bibitemNoStop [0]{.\EOS\space}%
\providecommand \EOS [0]{\spacefactor3000\relax}%
\providecommand \BibitemShut  [1]{\csname bibitem#1\endcsname}%
\let\auto@bib@innerbib\@empty
\bibitem [{\citenamefont {Keldysh}(1965)}]{keldysh1964diagram}%
  \BibitemOpen
  \bibfield  {author} {\bibinfo {author} {\bibfnamefont {L.~V.}\ \bibnamefont
  {Keldysh}},\ }\bibfield  {title} {\bibinfo {title} {{Diagram Technique for
  Nonequilibrium Processes}},\ }\href
  {https://doi.org/10.1142/9789811279461\_0007} {\bibfield  {journal} {\bibinfo
   {journal} {Sov. Phys. JETP}\ }\textbf {\bibinfo {volume} {20}},\ \bibinfo
  {pages} {1018} (\bibinfo {year} {1965})}\BibitemShut {NoStop}%
\bibitem [{\citenamefont {Kadanoff}\ and\ \citenamefont
  {Baym}(1962)}]{kadanoff1962quantum}%
  \BibitemOpen
  \bibfield  {author} {\bibinfo {author} {\bibfnamefont {L.~P.}\ \bibnamefont
  {Kadanoff}}\ and\ \bibinfo {author} {\bibfnamefont {G.}~\bibnamefont
  {Baym}},\ }\href@noop {} {\emph {\bibinfo {title} {Quantum Statistical
  Mechanics}}}\ (\bibinfo  {publisher} {W. A. Benjamin},\ \bibinfo {address}
  {New York},\ \bibinfo {year} {1962})\BibitemShut {NoStop}%
\bibitem [{\citenamefont {Danielewicz}(1984)}]{danielewicz1984quantum}%
  \BibitemOpen
  \bibfield  {author} {\bibinfo {author} {\bibfnamefont {P.}~\bibnamefont
  {Danielewicz}},\ }\bibfield  {title} {\bibinfo {title} {Quantum theory of
  nonequilibrium processes, i},\ }\href
  {https://doi.org/https://doi.org/10.1016/0003-4916(84)90092-7} {\bibfield
  {journal} {\bibinfo  {journal} {Ann. Phys. (N. Y.)}\ }\textbf {\bibinfo
  {volume} {152}},\ \bibinfo {pages} {239} (\bibinfo {year}
  {1984})}\BibitemShut {NoStop}%
\bibitem [{\citenamefont {Stefanucci}\ and\ \citenamefont {{van
  Leeuwen}}(2025)}]{stefanucci2013nonequilibrium}%
  \BibitemOpen
  \bibfield  {author} {\bibinfo {author} {\bibfnamefont {G.}~\bibnamefont
  {Stefanucci}}\ and\ \bibinfo {author} {\bibfnamefont {R.}~\bibnamefont {{van
  Leeuwen}}},\ }\href@noop {} {\emph {\bibinfo {title} {Nonequilibrium
  Many-Body Theory of Quantum Systems: A Modern Introduction}}},\ \bibinfo
  {edition} {2nd}\ ed.\ (\bibinfo  {publisher} {Cambridge University Press},\
  \bibinfo {year} {2025})\BibitemShut {NoStop}%
\bibitem [{\citenamefont {Balzer}\ and\ \citenamefont
  {Bonitz}(2013)}]{balzer2013nonequilibrium}%
  \BibitemOpen
  \bibfield  {author} {\bibinfo {author} {\bibfnamefont {K.}~\bibnamefont
  {Balzer}}\ and\ \bibinfo {author} {\bibfnamefont {M.}~\bibnamefont
  {Bonitz}},\ }\href {https://doi.org/10.1007/978-3-642-35082-5} {\emph
  {\bibinfo {title} {Nonequilibrium Green’s Functions Approach to
  Inhomogeneous Systems}}}\ (\bibinfo  {publisher} {Springer Berlin
  Heidelberg},\ \bibinfo {year} {2013})\BibitemShut {NoStop}%
\bibitem [{\citenamefont {Lipavsk\'y}\ \emph {et~al.}(1986)\citenamefont
  {Lipavsk\'y}, \citenamefont {Špička},\ and\ \citenamefont
  {Velick\'y}}]{lipavsky1986generalized}%
  \BibitemOpen
  \bibfield  {author} {\bibinfo {author} {\bibfnamefont {P.}~\bibnamefont
  {Lipavsk\'y}}, \bibinfo {author} {\bibfnamefont {V.}~\bibnamefont
  {Špička}},\ and\ \bibinfo {author} {\bibfnamefont {B.}~\bibnamefont
  {Velick\'y}},\ }\bibfield  {title} {\bibinfo {title} {Generalized
  kadanoff-baym ansatz for deriving quantum transport equations},\ }\href
  {https://doi.org/10.1103/PhysRevB.34.6933} {\bibfield  {journal} {\bibinfo
  {journal} {Phys. Rev. B}\ }\textbf {\bibinfo {volume} {34}},\ \bibinfo
  {pages} {6933} (\bibinfo {year} {1986})}\BibitemShut {NoStop}%
\bibitem [{\citenamefont {Schl\"unzen}\ \emph {et~al.}(2020)\citenamefont
  {Schl\"unzen}, \citenamefont {Joost},\ and\ \citenamefont
  {Bonitz}}]{schluenzen2020achieving}%
  \BibitemOpen
  \bibfield  {author} {\bibinfo {author} {\bibfnamefont {N.}~\bibnamefont
  {Schl\"unzen}}, \bibinfo {author} {\bibfnamefont {J.-P.}\ \bibnamefont
  {Joost}},\ and\ \bibinfo {author} {\bibfnamefont {M.}~\bibnamefont
  {Bonitz}},\ }\bibfield  {title} {\bibinfo {title} {Achieving the scaling
  limit for nonequilibrium green functions simulations},\ }\href
  {https://doi.org/10.1103/PhysRevLett.124.076601} {\bibfield  {journal}
  {\bibinfo  {journal} {Phys. Rev. Lett.}\ }\textbf {\bibinfo {volume} {124}},\
  \bibinfo {pages} {076601} (\bibinfo {year} {2020})}\BibitemShut {NoStop}%
\bibitem [{\citenamefont {Karlsson}\ \emph {et~al.}(2021)\citenamefont
  {Karlsson}, \citenamefont {{van Leeuwen}}, \citenamefont {Pavlyukh},
  \citenamefont {Perfetto},\ and\ \citenamefont
  {Stefanucci}}]{karlsson2021fast}%
  \BibitemOpen
  \bibfield  {author} {\bibinfo {author} {\bibfnamefont {D.}~\bibnamefont
  {Karlsson}}, \bibinfo {author} {\bibfnamefont {R.}~\bibnamefont {{van
  Leeuwen}}}, \bibinfo {author} {\bibfnamefont {Y.}~\bibnamefont {Pavlyukh}},
  \bibinfo {author} {\bibfnamefont {E.}~\bibnamefont {Perfetto}},\ and\
  \bibinfo {author} {\bibfnamefont {G.}~\bibnamefont {Stefanucci}},\ }\bibfield
   {title} {\bibinfo {title} {Fast green's function method for ultrafast
  electron-boson dynamics},\ }\href
  {https://doi.org/10.1103/PhysRevLett.127.036402} {\bibfield  {journal}
  {\bibinfo  {journal} {Phys. Rev. Lett.}\ }\textbf {\bibinfo {volume} {127}},\
  \bibinfo {pages} {036402} (\bibinfo {year} {2021})}\BibitemShut {NoStop}%
\bibitem [{\citenamefont {Pavlyukh}\ \emph
  {et~al.}(2022{\natexlab{a}})\citenamefont {Pavlyukh}, \citenamefont
  {Perfetto}, \citenamefont {Karlsson}, \citenamefont {{van Leeuwen}},\ and\
  \citenamefont {Stefanucci}}]{pavlyukh2022time-i}%
  \BibitemOpen
  \bibfield  {author} {\bibinfo {author} {\bibfnamefont {Y.}~\bibnamefont
  {Pavlyukh}}, \bibinfo {author} {\bibfnamefont {E.}~\bibnamefont {Perfetto}},
  \bibinfo {author} {\bibfnamefont {D.}~\bibnamefont {Karlsson}}, \bibinfo
  {author} {\bibfnamefont {R.}~\bibnamefont {{van Leeuwen}}},\ and\ \bibinfo
  {author} {\bibfnamefont {G.}~\bibnamefont {Stefanucci}},\ }\bibfield  {title}
  {\bibinfo {title} {Time-linear scaling nonequilibrium green's function
  methods for real-time simulations of interacting electrons and bosons. i.
  formalism},\ }\href {https://doi.org/10.1103/PhysRevB.105.125134} {\bibfield
  {journal} {\bibinfo  {journal} {Phys. Rev. B}\ }\textbf {\bibinfo {volume}
  {105}},\ \bibinfo {pages} {125134} (\bibinfo {year}
  {2022}{\natexlab{a}})}\BibitemShut {NoStop}%
\bibitem [{\citenamefont {Pavlyukh}\ \emph
  {et~al.}(2022{\natexlab{b}})\citenamefont {Pavlyukh}, \citenamefont
  {Perfetto}, \citenamefont {Karlsson}, \citenamefont {{van Leeuwen}},\ and\
  \citenamefont {Stefanucci}}]{pavlyukh2022time-ii}%
  \BibitemOpen
  \bibfield  {author} {\bibinfo {author} {\bibfnamefont {Y.}~\bibnamefont
  {Pavlyukh}}, \bibinfo {author} {\bibfnamefont {E.}~\bibnamefont {Perfetto}},
  \bibinfo {author} {\bibfnamefont {D.}~\bibnamefont {Karlsson}}, \bibinfo
  {author} {\bibfnamefont {R.}~\bibnamefont {{van Leeuwen}}},\ and\ \bibinfo
  {author} {\bibfnamefont {G.}~\bibnamefont {Stefanucci}},\ }\bibfield  {title}
  {\bibinfo {title} {Time-linear scaling nonequilibrium green's function method
  for real-time simulations of interacting electrons and bosons. ii. dynamics
  of polarons and doublons},\ }\href
  {https://doi.org/10.1103/PhysRevB.105.125135} {\bibfield  {journal} {\bibinfo
   {journal} {Phys. Rev. B}\ }\textbf {\bibinfo {volume} {105}},\ \bibinfo
  {pages} {125135} (\bibinfo {year} {2022}{\natexlab{b}})}\BibitemShut
  {NoStop}%
\bibitem [{\citenamefont {Balzer}\ \emph {et~al.}(2023)\citenamefont {Balzer},
  \citenamefont {Schl\"unzen}, \citenamefont {Ohldag}, \citenamefont {Joost},\
  and\ \citenamefont {Bonitz}}]{balzer2023accelerating}%
  \BibitemOpen
  \bibfield  {author} {\bibinfo {author} {\bibfnamefont {K.}~\bibnamefont
  {Balzer}}, \bibinfo {author} {\bibfnamefont {N.}~\bibnamefont {Schl\"unzen}},
  \bibinfo {author} {\bibfnamefont {H.}~\bibnamefont {Ohldag}}, \bibinfo
  {author} {\bibfnamefont {J.-P.}\ \bibnamefont {Joost}},\ and\ \bibinfo
  {author} {\bibfnamefont {M.}~\bibnamefont {Bonitz}},\ }\bibfield  {title}
  {\bibinfo {title} {Accelerating nonequilibrium green function simulations
  with embedding self-energies},\ }\href
  {https://doi.org/10.1103/PhysRevB.107.155141} {\bibfield  {journal} {\bibinfo
   {journal} {Phys. Rev. B}\ }\textbf {\bibinfo {volume} {107}},\ \bibinfo
  {pages} {155141} (\bibinfo {year} {2023})}\BibitemShut {NoStop}%
\bibitem [{\citenamefont {Tuovinen}\ \emph {et~al.}(2023)\citenamefont
  {Tuovinen}, \citenamefont {Pavlyukh}, \citenamefont {Perfetto},\ and\
  \citenamefont {Stefanucci}}]{tuovinen2023time}%
  \BibitemOpen
  \bibfield  {author} {\bibinfo {author} {\bibfnamefont {R.}~\bibnamefont
  {Tuovinen}}, \bibinfo {author} {\bibfnamefont {Y.}~\bibnamefont {Pavlyukh}},
  \bibinfo {author} {\bibfnamefont {E.}~\bibnamefont {Perfetto}},\ and\
  \bibinfo {author} {\bibfnamefont {G.}~\bibnamefont {Stefanucci}},\ }\bibfield
   {title} {\bibinfo {title} {Time-linear quantum transport simulations with
  correlated nonequilibrium green's functions},\ }\href
  {https://doi.org/10.1103/PhysRevLett.130.246301} {\bibfield  {journal}
  {\bibinfo  {journal} {Phys. Rev. Lett.}\ }\textbf {\bibinfo {volume} {130}},\
  \bibinfo {pages} {246301} (\bibinfo {year} {2023})}\BibitemShut {NoStop}%
\bibitem [{\citenamefont {Hermanns}\ \emph {et~al.}(2014)\citenamefont
  {Hermanns}, \citenamefont {Schl\"unzen},\ and\ \citenamefont
  {Bonitz}}]{hermanns2014hubbard}%
  \BibitemOpen
  \bibfield  {author} {\bibinfo {author} {\bibfnamefont {S.}~\bibnamefont
  {Hermanns}}, \bibinfo {author} {\bibfnamefont {N.}~\bibnamefont
  {Schl\"unzen}},\ and\ \bibinfo {author} {\bibfnamefont {M.}~\bibnamefont
  {Bonitz}},\ }\bibfield  {title} {\bibinfo {title} {Hubbard nanoclusters far
  from equilibrium},\ }\href {https://doi.org/10.1103/PhysRevB.90.125111}
  {\bibfield  {journal} {\bibinfo  {journal} {Phys. Rev. B}\ }\textbf {\bibinfo
  {volume} {90}},\ \bibinfo {pages} {125111} (\bibinfo {year}
  {2014})}\BibitemShut {NoStop}%
\bibitem [{\citenamefont {Joost}\ \emph {et~al.}(2022)\citenamefont {Joost},
  \citenamefont {Schl\"unzen}, \citenamefont {Ohldag}, \citenamefont {Bonitz},
  \citenamefont {Lackner},\ and\ \citenamefont
  {Březinov\'a}}]{joost2022dynamically}%
  \BibitemOpen
  \bibfield  {author} {\bibinfo {author} {\bibfnamefont {J.-P.}\ \bibnamefont
  {Joost}}, \bibinfo {author} {\bibfnamefont {N.}~\bibnamefont {Schl\"unzen}},
  \bibinfo {author} {\bibfnamefont {H.}~\bibnamefont {Ohldag}}, \bibinfo
  {author} {\bibfnamefont {M.}~\bibnamefont {Bonitz}}, \bibinfo {author}
  {\bibfnamefont {F.}~\bibnamefont {Lackner}},\ and\ \bibinfo {author}
  {\bibfnamefont {I.}~\bibnamefont {Březinov\'a}},\ }\bibfield  {title}
  {\bibinfo {title} {Dynamically screened ladder approximation: Simultaneous
  treatment of strong electronic correlations and dynamical screening out of
  equilibrium},\ }\href {https://doi.org/10.1103/PhysRevB.105.165155}
  {\bibfield  {journal} {\bibinfo  {journal} {Phys. Rev. B}\ }\textbf {\bibinfo
  {volume} {105}},\ \bibinfo {pages} {165155} (\bibinfo {year}
  {2022})}\BibitemShut {NoStop}%
\bibitem [{\citenamefont {Stefanucci}\ and\ \citenamefont
  {Perfetto}(2024)}]{stefanucci2024semiconductor}%
  \BibitemOpen
  \bibfield  {author} {\bibinfo {author} {\bibfnamefont {G.}~\bibnamefont
  {Stefanucci}}\ and\ \bibinfo {author} {\bibfnamefont {E.}~\bibnamefont
  {Perfetto}},\ }\bibfield  {title} {\bibinfo {title} {{Semiconductor
  electron-phonon equations: A rung above Boltzmann in the many-body ladder}},\
  }\href {https://doi.org/10.21468/SciPostPhys.16.3.073} {\bibfield  {journal}
  {\bibinfo  {journal} {SciPost Phys.}\ }\textbf {\bibinfo {volume} {16}},\
  \bibinfo {pages} {073} (\bibinfo {year} {2024})}\BibitemShut {NoStop}%
\bibitem [{\citenamefont {Kalvová}\ \emph {et~al.}(2019)\citenamefont
  {Kalvová}, \citenamefont {Velický},\ and\ \citenamefont
  {Špička}}]{kalvova2019beyond}%
  \BibitemOpen
  \bibfield  {author} {\bibinfo {author} {\bibfnamefont {A.}~\bibnamefont
  {Kalvová}}, \bibinfo {author} {\bibfnamefont {B.}~\bibnamefont {Velický}},\
  and\ \bibinfo {author} {\bibfnamefont {V.}~\bibnamefont {Špička}},\
  }\bibfield  {title} {\bibinfo {title} {Beyond the generalized kadanoff–baym
  ansatz},\ }\href {https://doi.org/https://doi.org/10.1002/pssb.201800594}
  {\bibfield  {journal} {\bibinfo  {journal} {Phys. Status Solidi B}\ }\textbf
  {\bibinfo {volume} {256}},\ \bibinfo {pages} {1800594} (\bibinfo {year}
  {2019})}\BibitemShut {NoStop}%
\bibitem [{\citenamefont {Kalvová}\ \emph {et~al.}(2023)\citenamefont
  {Kalvová}, \citenamefont {Špička}, \citenamefont {Velický},\ and\
  \citenamefont {Lipavský}}]{kalvova2023dynamical}%
  \BibitemOpen
  \bibfield  {author} {\bibinfo {author} {\bibfnamefont {A.}~\bibnamefont
  {Kalvová}}, \bibinfo {author} {\bibfnamefont {V.}~\bibnamefont {Špička}},
  \bibinfo {author} {\bibfnamefont {B.}~\bibnamefont {Velický}},\ and\
  \bibinfo {author} {\bibfnamefont {P.}~\bibnamefont {Lipavský}},\ }\bibfield
  {title} {\bibinfo {title} {Dynamical vertex correction to the generalized
  kadanoff-baym ansatz},\ }\href {https://doi.org/10.1209/0295-5075/acad9b}
  {\bibfield  {journal} {\bibinfo  {journal} {Europhys. Lett.}\ }\textbf
  {\bibinfo {volume} {141}},\ \bibinfo {pages} {16002} (\bibinfo {year}
  {2023})}\BibitemShut {NoStop}%
\bibitem [{\citenamefont {Pavlyukh}\ and\ \citenamefont
  {Tuovinen}(2025)}]{pavlyukh2025open}%
  \BibitemOpen
  \bibfield  {author} {\bibinfo {author} {\bibfnamefont {Y.}~\bibnamefont
  {Pavlyukh}}\ and\ \bibinfo {author} {\bibfnamefont {R.}~\bibnamefont
  {Tuovinen}},\ }\bibfield  {title} {\bibinfo {title} {Open system dynamics in
  linear time beyond the wide-band limit},\ }\href
  {https://doi.org/10.1103/PhysRevB.111.L241101} {\bibfield  {journal}
  {\bibinfo  {journal} {Phys. Rev. B}\ }\textbf {\bibinfo {volume} {111}},\
  \bibinfo {pages} {L241101} (\bibinfo {year} {2025})}\BibitemShut {NoStop}%
\bibitem [{\citenamefont {Tuovinen}\ and\ \citenamefont
  {Pavlyukh}(2025)}]{tuovinen2025thermoelectric}%
  \BibitemOpen
  \bibfield  {author} {\bibinfo {author} {\bibfnamefont {R.}~\bibnamefont
  {Tuovinen}}\ and\ \bibinfo {author} {\bibfnamefont {Y.}~\bibnamefont
  {Pavlyukh}},\ }\bibfield  {title} {\bibinfo {title} {Thermoelectric energy
  conversion in molecular junctions out of equilibrium},\ }\href
  {https://doi.org/10.1103/rj3h-8z3g} {\bibfield  {journal} {\bibinfo
  {journal} {PRX Energy}\ }\textbf {\bibinfo {volume} {4}},\ \bibinfo {pages}
  {043003} (\bibinfo {year} {2025})}\BibitemShut {NoStop}%
\bibitem [{\citenamefont {Säkkinen}\ \emph
  {et~al.}(2015{\natexlab{a}})\citenamefont {Säkkinen}, \citenamefont {Peng},
  \citenamefont {Appel},\ and\ \citenamefont {{van
  Leeuwen}}}]{sakkinen2015many}%
  \BibitemOpen
  \bibfield  {author} {\bibinfo {author} {\bibfnamefont {N.}~\bibnamefont
  {Säkkinen}}, \bibinfo {author} {\bibfnamefont {Y.}~\bibnamefont {Peng}},
  \bibinfo {author} {\bibfnamefont {H.}~\bibnamefont {Appel}},\ and\ \bibinfo
  {author} {\bibfnamefont {R.}~\bibnamefont {{van Leeuwen}}},\ }\bibfield
  {title} {\bibinfo {title} {Many-body green’s function theory for
  electron-phonon interactions: Ground state properties of the holstein
  dimer},\ }\href {https://doi.org/10.1063/1.4936142} {\bibfield  {journal}
  {\bibinfo  {journal} {J. Chem. Phys.}\ }\textbf {\bibinfo {volume} {143}},\
  \bibinfo {pages} {234101} (\bibinfo {year} {2015}{\natexlab{a}})}\BibitemShut
  {NoStop}%
\bibitem [{\citenamefont {Säkkinen}\ \emph
  {et~al.}(2015{\natexlab{b}})\citenamefont {Säkkinen}, \citenamefont {Peng},
  \citenamefont {Appel},\ and\ \citenamefont {{van
  Leeuwen}}}]{sakkinen2015many-ii}%
  \BibitemOpen
  \bibfield  {author} {\bibinfo {author} {\bibfnamefont {N.}~\bibnamefont
  {Säkkinen}}, \bibinfo {author} {\bibfnamefont {Y.}~\bibnamefont {Peng}},
  \bibinfo {author} {\bibfnamefont {H.}~\bibnamefont {Appel}},\ and\ \bibinfo
  {author} {\bibfnamefont {R.}~\bibnamefont {{van Leeuwen}}},\ }\bibfield
  {title} {\bibinfo {title} {Many-body green’s function theory for
  electron-phonon interactions: The kadanoff-baym approach to spectral
  properties of the holstein dimer},\ }\href
  {https://doi.org/10.1063/1.4936143} {\bibfield  {journal} {\bibinfo
  {journal} {J. Chem. Phys.}\ }\textbf {\bibinfo {volume} {143}},\ \bibinfo
  {pages} {234102} (\bibinfo {year} {2015}{\natexlab{b}})}\BibitemShut
  {NoStop}%
\bibitem [{\citenamefont {Bonitz}\ \emph {et~al.}(1996)\citenamefont {Bonitz},
  \citenamefont {Kremp}, \citenamefont {Scott}, \citenamefont {Binder},
  \citenamefont {Kraeft},\ and\ \citenamefont {Köhler}}]{bonitz1996numerical}%
  \BibitemOpen
  \bibfield  {author} {\bibinfo {author} {\bibfnamefont {M.}~\bibnamefont
  {Bonitz}}, \bibinfo {author} {\bibfnamefont {D.}~\bibnamefont {Kremp}},
  \bibinfo {author} {\bibfnamefont {D.~C.}\ \bibnamefont {Scott}}, \bibinfo
  {author} {\bibfnamefont {R.}~\bibnamefont {Binder}}, \bibinfo {author}
  {\bibfnamefont {W.~D.}\ \bibnamefont {Kraeft}},\ and\ \bibinfo {author}
  {\bibfnamefont {H.~S.}\ \bibnamefont {Köhler}},\ }\bibfield  {title}
  {\bibinfo {title} {Numerical analysis of non-markovian effects in
  charge-carrier scattering: one-time versus two-time kinetic equations},\
  }\href {https://doi.org/10.1088/0953-8984/8/33/012} {\bibfield  {journal}
  {\bibinfo  {journal} {J. Phys. Condens. Matter}\ }\textbf {\bibinfo {volume}
  {8}},\ \bibinfo {pages} {6057} (\bibinfo {year} {1996})}\BibitemShut
  {NoStop}%
\bibitem [{\citenamefont {Jahnke}\ \emph {et~al.}(1997)\citenamefont {Jahnke},
  \citenamefont {Kira},\ and\ \citenamefont {Koch}}]{jahnke1997linear}%
  \BibitemOpen
  \bibfield  {author} {\bibinfo {author} {\bibfnamefont {F.}~\bibnamefont
  {Jahnke}}, \bibinfo {author} {\bibfnamefont {M.}~\bibnamefont {Kira}},\ and\
  \bibinfo {author} {\bibfnamefont {S.~W.}\ \bibnamefont {Koch}},\ }\bibfield
  {title} {\bibinfo {title} {Linear and nonlinear optical properties of
  excitons in semiconductor quantum wells and microcavities},\ }\href
  {https://doi.org/10.1007/s002570050490} {\bibfield  {journal} {\bibinfo
  {journal} {Z. Phys. B}\ }\textbf {\bibinfo {volume} {104}},\ \bibinfo {pages}
  {559} (\bibinfo {year} {1997})}\BibitemShut {NoStop}%
\bibitem [{\citenamefont {Pal}\ \emph {et~al.}(2009)\citenamefont {Pal},
  \citenamefont {Pavlyukh}, \citenamefont {Schneider},\ and\ \citenamefont
  {H{\"u}bner}}]{pal2009conserving}%
  \BibitemOpen
  \bibfield  {author} {\bibinfo {author} {\bibfnamefont {G.}~\bibnamefont
  {Pal}}, \bibinfo {author} {\bibfnamefont {Y.}~\bibnamefont {Pavlyukh}},
  \bibinfo {author} {\bibfnamefont {H.~C.}\ \bibnamefont {Schneider}},\ and\
  \bibinfo {author} {\bibfnamefont {W.}~\bibnamefont {H{\"u}bner}},\ }\bibfield
   {title} {\bibinfo {title} {Conserving quasiparticle calculations for small
  metal clusters},\ }\href {https://doi.org/10.1140/epjb/e2009-00253-9}
  {\bibfield  {journal} {\bibinfo  {journal} {Eur. Phys. J. B}\ }\textbf
  {\bibinfo {volume} {70}},\ \bibinfo {pages} {483} (\bibinfo {year}
  {2009})}\BibitemShut {NoStop}%
\bibitem [{\citenamefont {Makait}\ \emph {et~al.}(2023)\citenamefont {Makait},
  \citenamefont {Fajardo},\ and\ \citenamefont {Bonitz}}]{makait2023time}%
  \BibitemOpen
  \bibfield  {author} {\bibinfo {author} {\bibfnamefont {C.}~\bibnamefont
  {Makait}}, \bibinfo {author} {\bibfnamefont {F.~B.}\ \bibnamefont
  {Fajardo}},\ and\ \bibinfo {author} {\bibfnamefont {M.}~\bibnamefont
  {Bonitz}},\ }\bibfield  {title} {\bibinfo {title} {Time-dependent charged
  particle stopping in quantum plasmas: testing the g1–g2 scheme for
  quasi-one-dimensional systems},\ }\href
  {https://doi.org/https://doi.org/10.1002/ctpp.202300008} {\bibfield
  {journal} {\bibinfo  {journal} {Contrib. Plasma Phys.}\ }\textbf {\bibinfo
  {volume} {63}},\ \bibinfo {pages} {e202300008} (\bibinfo {year}
  {2023})}\BibitemShut {NoStop}%
\bibitem [{\citenamefont {Latini}\ \emph {et~al.}(2014)\citenamefont {Latini},
  \citenamefont {Perfetto}, \citenamefont {Uimonen}, \citenamefont {{van
  Leeuwen}},\ and\ \citenamefont {Stefanucci}}]{latini2014charge}%
  \BibitemOpen
  \bibfield  {author} {\bibinfo {author} {\bibfnamefont {S.}~\bibnamefont
  {Latini}}, \bibinfo {author} {\bibfnamefont {E.}~\bibnamefont {Perfetto}},
  \bibinfo {author} {\bibfnamefont {A.-M.}\ \bibnamefont {Uimonen}}, \bibinfo
  {author} {\bibfnamefont {R.}~\bibnamefont {{van Leeuwen}}},\ and\ \bibinfo
  {author} {\bibfnamefont {G.}~\bibnamefont {Stefanucci}},\ }\bibfield  {title}
  {\bibinfo {title} {Charge dynamics in molecular junctions: Nonequilibrium
  green's function approach made fast},\ }\href
  {https://doi.org/10.1103/PhysRevB.89.075306} {\bibfield  {journal} {\bibinfo
  {journal} {Phys. Rev. B}\ }\textbf {\bibinfo {volume} {89}},\ \bibinfo
  {pages} {075306} (\bibinfo {year} {2014})}\BibitemShut {NoStop}%
\bibitem [{\citenamefont {Bonitz}\ \emph {et~al.}(2019)\citenamefont {Bonitz},
  \citenamefont {Balzer}, \citenamefont {Schlünzen}, \citenamefont
  {Rasmussen},\ and\ \citenamefont {Joost}}]{bonitz2019ion}%
  \BibitemOpen
  \bibfield  {author} {\bibinfo {author} {\bibfnamefont {M.}~\bibnamefont
  {Bonitz}}, \bibinfo {author} {\bibfnamefont {K.}~\bibnamefont {Balzer}},
  \bibinfo {author} {\bibfnamefont {N.}~\bibnamefont {Schlünzen}}, \bibinfo
  {author} {\bibfnamefont {M.~R.}\ \bibnamefont {Rasmussen}},\ and\ \bibinfo
  {author} {\bibfnamefont {J.-P.}\ \bibnamefont {Joost}},\ }\bibfield  {title}
  {\bibinfo {title} {Ion impact induced ultrafast electron dynamics in finite
  graphene-type hubbard clusters},\ }\href
  {https://doi.org/https://doi.org/10.1002/pssb.201800490} {\bibfield
  {journal} {\bibinfo  {journal} {Phys. Status Solidi B}\ }\textbf {\bibinfo
  {volume} {256}},\ \bibinfo {pages} {1800490} (\bibinfo {year}
  {2019})}\BibitemShut {NoStop}%
\bibitem [{\citenamefont {Keating}(1968)}]{keating1968dielectric}%
  \BibitemOpen
  \bibfield  {author} {\bibinfo {author} {\bibfnamefont {P.~N.}\ \bibnamefont
  {Keating}},\ }\bibfield  {title} {\bibinfo {title} {Dielectric screening and
  the phonon spectra of metallic and nonmetallic crystals},\ }\href
  {https://doi.org/10.1103/PhysRev.175.1171} {\bibfield  {journal} {\bibinfo
  {journal} {Phys. Rev.}\ }\textbf {\bibinfo {volume} {175}},\ \bibinfo {pages}
  {1171} (\bibinfo {year} {1968})}\BibitemShut {NoStop}%
\bibitem [{\citenamefont {Vogl}(1976)}]{vogl1976microscopic}%
  \BibitemOpen
  \bibfield  {author} {\bibinfo {author} {\bibfnamefont {P.}~\bibnamefont
  {Vogl}},\ }\bibfield  {title} {\bibinfo {title} {Microscopic theory of
  electron-phonon interaction in insulators or semiconductors},\ }\href
  {https://doi.org/10.1103/PhysRevB.13.694} {\bibfield  {journal} {\bibinfo
  {journal} {Phys. Rev. B}\ }\textbf {\bibinfo {volume} {13}},\ \bibinfo
  {pages} {694} (\bibinfo {year} {1976})}\BibitemShut {NoStop}%
\bibitem [{\citenamefont {Giustino}(2017)}]{giustino2017electron}%
  \BibitemOpen
  \bibfield  {author} {\bibinfo {author} {\bibfnamefont {F.}~\bibnamefont
  {Giustino}},\ }\bibfield  {title} {\bibinfo {title} {Electron-phonon
  interactions from first principles},\ }\href
  {https://doi.org/10.1103/RevModPhys.89.015003} {\bibfield  {journal}
  {\bibinfo  {journal} {Rev. Mod. Phys.}\ }\textbf {\bibinfo {volume} {89}},\
  \bibinfo {pages} {015003} (\bibinfo {year} {2017})}\BibitemShut {NoStop}%
\bibitem [{\citenamefont {Hu}\ \emph {et~al.}(2010)\citenamefont {Hu},
  \citenamefont {Xu},\ and\ \citenamefont {Yan}}]{hu2010communication}%
  \BibitemOpen
  \bibfield  {author} {\bibinfo {author} {\bibfnamefont {J.}~\bibnamefont
  {Hu}}, \bibinfo {author} {\bibfnamefont {R.-X.}\ \bibnamefont {Xu}},\ and\
  \bibinfo {author} {\bibfnamefont {Y.}~\bibnamefont {Yan}},\ }\bibfield
  {title} {\bibinfo {title} {Communication: Padé spectrum decomposition of
  fermi function and bose function},\ }\href
  {https://doi.org/10.1063/1.3484491} {\bibfield  {journal} {\bibinfo
  {journal} {J. Chem. Phys.}\ }\textbf {\bibinfo {volume} {133}},\ \bibinfo
  {pages} {101106} (\bibinfo {year} {2010})}\BibitemShut {NoStop}%
\bibitem [{\citenamefont {Pavlyukh}\ \emph {et~al.}(2024)\citenamefont
  {Pavlyukh}, \citenamefont {Tuovinen}, \citenamefont {Perfetto},\ and\
  \citenamefont {Stefanucci}}]{pavlyukh2025cheers}%
  \BibitemOpen
  \bibfield  {author} {\bibinfo {author} {\bibfnamefont {Y.}~\bibnamefont
  {Pavlyukh}}, \bibinfo {author} {\bibfnamefont {R.}~\bibnamefont {Tuovinen}},
  \bibinfo {author} {\bibfnamefont {E.}~\bibnamefont {Perfetto}},\ and\
  \bibinfo {author} {\bibfnamefont {G.}~\bibnamefont {Stefanucci}},\ }\bibfield
   {title} {\bibinfo {title} {Cheers: A linear-scaling kbe+gkba code},\ }\href
  {https://doi.org/https://doi.org/10.1002/pssb.202300504} {\bibfield
  {journal} {\bibinfo  {journal} {Phys. Status Solidi B}\ }\textbf {\bibinfo
  {volume} {261}},\ \bibinfo {pages} {2300504} (\bibinfo {year}
  {2024})}\BibitemShut {NoStop}%
\bibitem [{\citenamefont {Pavlyukh}(2024)}]{pavlyukh2024nonequilibrium}%
  \BibitemOpen
  \bibfield  {author} {\bibinfo {author} {\bibfnamefont {Y.}~\bibnamefont
  {Pavlyukh}},\ }\bibfield  {title} {\bibinfo {title} {Nonequilibrium dynamics
  of the hubbard dimer},\ }\href
  {https://doi.org/https://doi.org/10.1002/pssb.202300510} {\bibfield
  {journal} {\bibinfo  {journal} {Phys. Status Solidi B}\ }\textbf {\bibinfo
  {volume} {261}},\ \bibinfo {pages} {2300510} (\bibinfo {year} {2024})},\
  \Eprint
  {https://arxiv.org/abs/https://onlinelibrary.wiley.com/doi/pdf/10.1002/pssb.202300510}
  {https://onlinelibrary.wiley.com/doi/pdf/10.1002/pssb.202300510} \BibitemShut
  {NoStop}%
\bibitem [{\citenamefont {Tuovinen}\ \emph {et~al.}(2020)\citenamefont
  {Tuovinen}, \citenamefont {Golež}, \citenamefont {Eckstein},\ and\
  \citenamefont {Sentef}}]{tuovinen2020comparing}%
  \BibitemOpen
  \bibfield  {author} {\bibinfo {author} {\bibfnamefont {R.}~\bibnamefont
  {Tuovinen}}, \bibinfo {author} {\bibfnamefont {D.}~\bibnamefont {Golež}},
  \bibinfo {author} {\bibfnamefont {M.}~\bibnamefont {Eckstein}},\ and\
  \bibinfo {author} {\bibfnamefont {M.~A.}\ \bibnamefont {Sentef}},\ }\bibfield
   {title} {\bibinfo {title} {Comparing the generalized kadanoff-baym ansatz
  with the full kadanoff-baym equations for an excitonic insulator out of
  equilibrium},\ }\href {https://doi.org/10.1103/PhysRevB.102.115157}
  {\bibfield  {journal} {\bibinfo  {journal} {Phys. Rev. B}\ }\textbf {\bibinfo
  {volume} {102}},\ \bibinfo {pages} {115157} (\bibinfo {year}
  {2020})}\BibitemShut {NoStop}%
\bibitem [{\citenamefont {Tuovinen}\ \emph {et~al.}(2021)\citenamefont
  {Tuovinen}, \citenamefont {{van Leeuwen}}, \citenamefont {Perfetto},\ and\
  \citenamefont {Stefanucci}}]{tuovinen2021electronic}%
  \BibitemOpen
  \bibfield  {author} {\bibinfo {author} {\bibfnamefont {R.}~\bibnamefont
  {Tuovinen}}, \bibinfo {author} {\bibfnamefont {R.}~\bibnamefont {{van
  Leeuwen}}}, \bibinfo {author} {\bibfnamefont {E.}~\bibnamefont {Perfetto}},\
  and\ \bibinfo {author} {\bibfnamefont {G.}~\bibnamefont {Stefanucci}},\
  }\bibfield  {title} {\bibinfo {title} {Electronic transport in molecular
  junctions: The generalized kadanoff–baym ansatz with initial contact and
  correlations},\ }\href {https://doi.org/10.1063/5.0040685} {\bibfield
  {journal} {\bibinfo  {journal} {J. Chem. Phys.}\ }\textbf {\bibinfo {volume}
  {154}},\ \bibinfo {pages} {094104} (\bibinfo {year} {2021})}\BibitemShut
  {NoStop}%
\bibitem [{\citenamefont {Albrecht}\ \emph {et~al.}(2013)\citenamefont
  {Albrecht}, \citenamefont {Martin-Rodero}, \citenamefont {Monreal},
  \citenamefont {M\"uhlbacher},\ and\ \citenamefont
  {Levy~Yeyati}}]{albrecht2013long}%
  \BibitemOpen
  \bibfield  {author} {\bibinfo {author} {\bibfnamefont {K.~F.}\ \bibnamefont
  {Albrecht}}, \bibinfo {author} {\bibfnamefont {A.}~\bibnamefont
  {Martin-Rodero}}, \bibinfo {author} {\bibfnamefont {R.~C.}\ \bibnamefont
  {Monreal}}, \bibinfo {author} {\bibfnamefont {L.}~\bibnamefont
  {M\"uhlbacher}},\ and\ \bibinfo {author} {\bibfnamefont {A.}~\bibnamefont
  {Levy~Yeyati}},\ }\bibfield  {title} {\bibinfo {title} {Long transient
  dynamics in the anderson-holstein model out of equilibrium},\ }\href
  {https://doi.org/10.1103/PhysRevB.87.085127} {\bibfield  {journal} {\bibinfo
  {journal} {Phys. Rev. B}\ }\textbf {\bibinfo {volume} {87}},\ \bibinfo
  {pages} {085127} (\bibinfo {year} {2013})}\BibitemShut {NoStop}%
\bibitem [{\citenamefont {Seoane~Souto}\ \emph {et~al.}(2015)\citenamefont
  {Seoane~Souto}, \citenamefont {Avriller}, \citenamefont {Monreal},
  \citenamefont {Mart\'{\i}n-Rodero},\ and\ \citenamefont
  {Levy~Yeyati}}]{souto2015transient}%
  \BibitemOpen
  \bibfield  {author} {\bibinfo {author} {\bibfnamefont {R.}~\bibnamefont
  {Seoane~Souto}}, \bibinfo {author} {\bibfnamefont {R.}~\bibnamefont
  {Avriller}}, \bibinfo {author} {\bibfnamefont {R.~C.}\ \bibnamefont
  {Monreal}}, \bibinfo {author} {\bibfnamefont {A.}~\bibnamefont
  {Mart\'{\i}n-Rodero}},\ and\ \bibinfo {author} {\bibfnamefont
  {A.}~\bibnamefont {Levy~Yeyati}},\ }\bibfield  {title} {\bibinfo {title}
  {Transient dynamics and waiting time distribution of molecular junctions in
  the polaronic regime},\ }\href {https://doi.org/10.1103/PhysRevB.92.125435}
  {\bibfield  {journal} {\bibinfo  {journal} {Phys. Rev. B}\ }\textbf {\bibinfo
  {volume} {92}},\ \bibinfo {pages} {125435} (\bibinfo {year}
  {2015})}\BibitemShut {NoStop}%
\bibitem [{\citenamefont {Souto}\ \emph {et~al.}(2018)\citenamefont {Souto},
  \citenamefont {Avriller}, \citenamefont {Yeyati},\ and\ \citenamefont
  {Martín-Rodero}}]{souto2018transient}%
  \BibitemOpen
  \bibfield  {author} {\bibinfo {author} {\bibfnamefont {R.~S.}\ \bibnamefont
  {Souto}}, \bibinfo {author} {\bibfnamefont {R.}~\bibnamefont {Avriller}},
  \bibinfo {author} {\bibfnamefont {A.~L.}\ \bibnamefont {Yeyati}},\ and\
  \bibinfo {author} {\bibfnamefont {A.}~\bibnamefont {Martín-Rodero}},\
  }\bibfield  {title} {\bibinfo {title} {Transient dynamics in interacting
  nanojunctions within self-consistent perturbation theory},\ }\href
  {https://doi.org/10.1088/1367-2630/aad99d} {\bibfield  {journal} {\bibinfo
  {journal} {New J. Phys.}\ }\textbf {\bibinfo {volume} {20}},\ \bibinfo
  {pages} {083039} (\bibinfo {year} {2018})}\BibitemShut {NoStop}%
\end{thebibliography}

%

\end{document}